\documentclass[final,english]{bullsrsl}[2022/06/15]

\usepackage[latin1]{inputenc}
\usepackage[T1]{fontenc}

\usepackage{natbib} 
\usepackage{graphicx}

\begin{document}
\title{Measuring Hydrogen-to-Helium Ratio in Cool Stars}

\author[corresponding]{Banagere Prakash}{Hema}
\author{Gajendra}{Pandey}
\affiliation{Indian Institute of Astrophysics, Bengaluru, Karnataka 560034, India}
\correspondance{hema.bp@iiap.res.in, pandey@iiap.res.in}
\date{13th October 2020}
\maketitle



\begin{abstract}
Conventionally, the helium-to-hydrogen ratio for the stars 
are adopted to be 0.1, as standard, unless, the stars are severely
deficient in hydrogen like in RCB-class, or the stars' helium abundance 
is accurately measured using He\,{\sc i} transitions in warm/hotter
stars. In our study, the small change in helium-to-hydrogen ratio 
(from standard value, 0.1) in normal giants were detected from 
the large difference ($>$ 0.3 dex) in the Mg-abundance measured 
from Mg\,{\sc i} lines and the subordinate lines of (0,0) MgH band.  
These are the stars that are mildly hydrogen-deficient/He-enhanced.
Such stars were spectroscopically discovered for the first time  
among giants of the globular cluster $\omega$ Centauri. The sample 
selection, observations, methodoloy and results are discussed in detail.
\end{abstract}


\section{Introduction}
The universe is vast and mysterious that it always unfolds itself to 
new, unseen and unknown things hidden in it. The stars that are
studied are all abundant in hydrogen, except for a handful of stars 
that are severely deficient in hydrogen \citep{asplund00, lambert94, 
hema12a,hema12b,hema17, pandey06, pandey08, clayton05}. 
We conducted a survey for exploring the hydrogen abundance in giants
of the massive and brightest Galactic globular cluster (GGC) 
$\omega$ Centauri of our Galaxy, the Milky Way. 

The observed large spread in the metallicity ([Fe/H]) and the
other abundance anomalies of $\omega$ Cen cluster stars 
\citep{johnson10, marino11, simpson13, sollima05} 
including the existence of multiple stellar
populations, the He-normal and He-enhanced or H-poor 
\citep{bedin04, piotto05}, makes it an enigmatic GGC.
The recent spectroscopic studies of \citet{dupree13} and
\citet{marino14} confirm the existence of the He-enhanced
stars in $\omega$ Cen and NGC 2808, respectively. Hence, our survey
also explores the H-deficiency or the He-enhancement in the
sample giants of $\omega$ Cen, but spectroscopically.
The observations, methodology, the results and discussions are 
explained in the following sections.

\section{Sample Stars and Observations}

In order to investigate this large spread in the metallicity 
of $\omega$ Cen,  a survey was conducted among the 
brightest giants of the cluster. 
We suspected that, due to different hydrogen-to-helium ratio 
in these stars, the metallicities appear different.
Hence, we selected the brightest giants of different metallicities 
for our survey. 

Note that all the metal-rich giants (+0.5 $>$ [Fe/H] $>$ -0.5)
were selected for observations, irrespective of their IR-colors,
and totaled 130 in number. However, the metal-poor giants
(-0.5 $>$ [Fe/H] $>$ -2.5) selected for observations, with
IR-colors like RCB stars, were 40 in number. Though the sample
of red giants from the core of $\omega$  Cen were not included in
our sample (to avoid confusion in identifying the giants in the
crowded field), many of the giants in the periphery were double
or multiple objects. The giants that were not clearly resolved
were excluded from our observations. Hence, only 34 of the
130 metal-rich stars and about 11 of the 40 metal-poor stars
were selected for observations. These program stars (11 in number) 
analyzed for investigating their He-enhancement with their 
low-resolution spectra are given in Table\,\ref{tab:1}. 

\begin{table}
\centering
\caption{The stellar parameters, metallicities and
derived Mg abundances for the program stars
in the order of their increasing $T_{\rm eff}$. The stars in 
italics are the relatively H-poor/He-enhanced.
\label{tab:1}}
\bigskip
\begin{tabular}{cccccccccr}
\hline
\textbf{Star} & \textbf{Star}   & \textbf{S/N} & \textbf{\textit{T}$_\mathbf{eff}$} & \textbf{log\,\textit{g}} & \textbf{[Fe/H]} & \textbf{Group} & \textbf{log $\epsilon$(Mg)} &  \textbf{[Mg/Fe]} & \textbf{He/H} \\
              & \textbf{(LEID)} &              &                      &                   &                 &                &                             &                   &               \\
\hline
269309 &   ...   & 70 & 3760 & 0.90 & $-$0.5 & First & 7.1$\pm$0.2 & 0.0 & ... \\
{\it 73170}  & 39048 & 100 & 3965 & 0.95 & $-$0.65 & First & 6.75$\pm$0.2 & $-$0.2 & 0.20 \\
{\it 178243} & 60073 & 100 & 3985 & 0.75 & $-$0.8 & First & 6.4$\pm$0.2 & $-$0.4& ... \\
172980 & 61067 & 110 & 4035 & 0.85 & $-$1.0 & First & 7.0$\pm$0.2 & $+$0.4 & 0.1 \\
178691 & 50193 & 110 & 4075 & 0.65 & $-$1.2 & First & 6.6$\pm$0.2 & $+$0.2 & ... \\
271054 &   ...   & 100 & 4100 & 1.40 & $-$1.0 & First & 6.7$\pm$0.2 & $+$0.1 & ... \\
40867  & 54022 & 110 & 4135 & 1.15 & $-$0.5 & First & 7.2$\pm$0.2 & $+$0.1 & ... \\
250000 &   ...  & 90 & 4175 & 1.40 & $-$1.0 & First & 6.9$\pm$0.2 & $+$0.3 & ... \\
131105 & 51074 & 80 & 4180 & 1.05 & $-$1.1  & First & 6.9$\pm$0.2 & $+$0.4 & ... \\
166240 & 55101 & 60 & 4240 & 1.15 & $-$1.0 & First & 6.8$\pm$0.2 & $+$0.2 & ... \\
{\it 262788} & 34225 & 110 & 4265 & 1.30 & $-$1.0 & Third & $<$6.0$\pm$0.2 & $<-$0.6 & 0.15  \\
251701 & 32169 & 100 & 4285 & 1.35 & $-$1.0 & First & 7.0$\pm$0.2 & $+$0.4 & 0.1 \\
{\it 193804} & 35201 & 80 & 4335 & 1.10 & $-$1.0 & Third & $<$6.5$\pm$0.2 & $<-$0.1 & ... \\
5001638 &   ...   & 150 & 4400 & 1.6 & $-$0.5 & First & 7.3$\pm$0.2 & $+$0.2 & ... \\
270931 &   ...   & 100 & 4420 & 1.25 & $-$0.5 & First & 7.2$\pm$0.2 & $+$0.1 & ... \\
214247 & 37275 & 60 & 4430 & 1.45 & $-$1.5 & Third & 6.5$\pm$0.2 & $+$0.4 & ... \\
216815 & 43475 & 80 & 4500 & 1.85 & $-$0.6 & First & 7.3$\pm$0.2 & $+$0.3 & ...\\
14943 &  53012 & 100 &  4605 &  1.35 & $-$1.8 & Second & $<$6.7$\pm$0.2 &  $<+$0.9 & ...\\
\hline
\end{tabular}
\end{table}

Low-resolution optical spectra for these selected red giants
of $\omega$ Cen were obtained from the 2.34 m Vainu Bappu
Telescope (VBT), the Vainu Bappu Observatory, equipped with
the Optomechanics Research spectrograph (Prabhu et al. 1998)
and the 1 K $\times$ 1 K CCD camera. These spectra obtained using
600 l mm$^{-1}$ grating centered at the H$\alpha$ line 
at 6563\AA, were at a resolution of about 8\AA. 

From the low-resolution studies, about four He-enhanced 
stars were discovered. Out of these, two He-enhanced stars along 
with their normal comparison stars were analyzed by obtaining 
the high-resolution spectra. These four stars that are analyzed 
with their high-resolution spectra are given in italics in Table\,\ref{tab:1}.
The high-resolution optical spectra were obtained using the
Southern African Large Telescope (SALT) high-resolution
spectrograph (HRS). These spectra obtained with SALT-HRS
have a resolving power R ($\lambda/\Delta\lambda$) of 40,000. 
The spectra were obtained with both blue and red cameras 
using 2K $\times$ 4K and 4K $\times$ 4K CCDs, respectively, 
spanning a spectral range of 370 -- 550 nm in the blue and 
550 -- 890 nm in the red. The data reduction and analyses were
carried out using the IRAF software package.

\section{Methodology and Results}

\begin{figure}[t!]
\centering
\includegraphics[width=12cm]{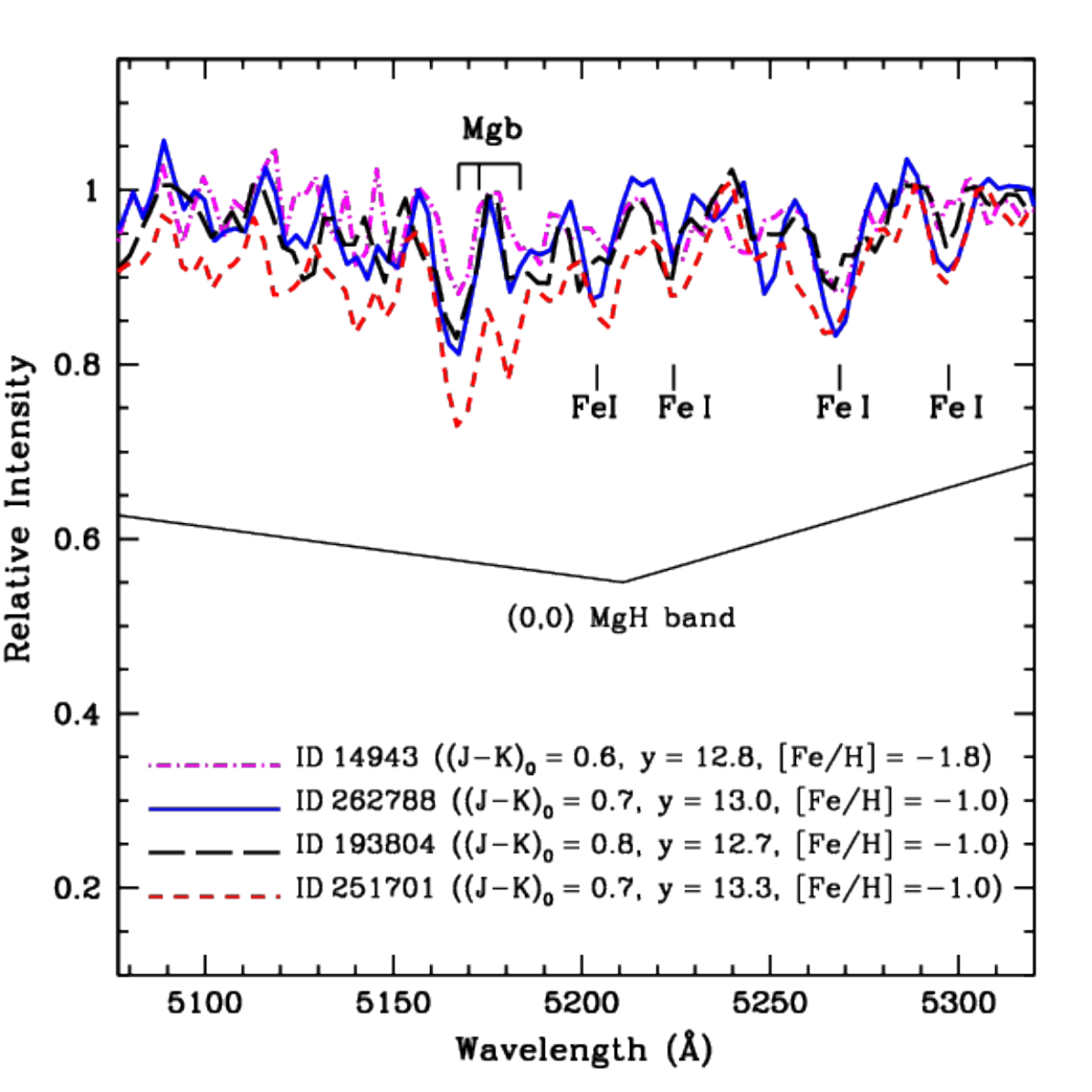}
\begin{minipage}{12cm}
\caption{The observed spectrum
of 262788 (in thick blue line) and 193804 (in black long dashed line),
the third group stars,
compared with the observed spectrum of 251701
(in red dashed line), the first group
star. Also shown is the observed spectrum of 14943
(in magenta dash dotted line), the second group
star. These stars have similar $(J-K)_{0}$ colour
and y magnitudes.
The key features such as Mg\,$b$ lines, the MgH band and
the Fe\,{\sc i} lines are marked.
The [Fe/H] values are from \citet{johnson10}.
\label{fig:1}
}
\end{minipage}
\end{figure}

The observed spectra of all the program stars were continuum
normalized. The region of the spectrum (having maximum flux)
free of absorption lines is treated as the continuum point, and
a smooth curve passing through these points is defined as the
continuum. The well defined continuum in the spectrum of the
sample metal-poor giant, and in the spectrum of Arcturus with
very high signal-to-noise ratio (S/N), is used as a reference
for judging the continuum for the sample metal-rich stars in the
wavelength window 4900--5400 \AA\, including the Mg b triplet and
the complete MgH band. The analyses of the observed spectra of
the program stars were carried out based on the strengths of the
blue degraded (0,0) MgH band extending from 5330 to 4950 \AA,
with the band head at 5211 \AA, and the Mg b lines at 5167.32 \AA,
5172.68 \AA, and 5183.60 \AA. 

Based on the strengths of these
features in the observed spectra, three groups were identified
in our sample: (1) the metal-rich giants with strong Mg b lines
and the MgH band, (2) the metal-poor giants with weak Mg b
lines and no MgH band, and (3) the metal-rich giants with strong
Mg b lines, but no MgH band (see Fig.\,\ref{fig:1}). 
To analyze the strengths of the
MgH band in the observed spectra of sample stars, the stars with
similar $(J-K)_{0}$ colors ($\Delta(J-K)_{0}$ $\sim$ $\pm$ 0.1), 
and y magnitudes ($\Delta$y $\sim$ $\pm$ 0.5) that 
represent the effective temperatures ($T_{\rm eff}$) and
surface gravities (log g), respectively, were selected. The spectra
of stars having similar y magnitude and $(J-K)_{0}$ colors were
then compared with each other. From this comparison, four
stars were identified having a weaker or absent MgH band than
expected. Two stars, 178243 and 73170, are from the first group
showing the strong Mg b lines, but a weaker MgH band than
expected for their stellar parameters (see Fig.\,1 of \citet{hema14}). 
For all the program stars listed in Table\,\ref{tab:1}, the stellar parameters 
and the metallicities derived from the 
high-resolution spectra were adopted from \citet{johnson10}.

The other two stars, 262788 and 193804, are from the third group showing
relatively strong Mg b lines, but an absent MgH band which was
unexpected for their stellar parameters. Judging
by the observed strengths of the Mg b lines and the presence/
absence of the MgH band expected for their stellar parameters,
the spectra of the giants 178243, 73170, 262788, and 193804
suggest that their atmospheres are relatively H-poor.
Hence, to confirm this suggestion, the observed strengths of the MgH
bands were further analyzed by synthesizing the spectra of these
four stars along with the program stars of the first, second,
and third groups for their adopted stellar parameters.

\begin{figure}[t!]
\centering
\includegraphics[width=12cm]{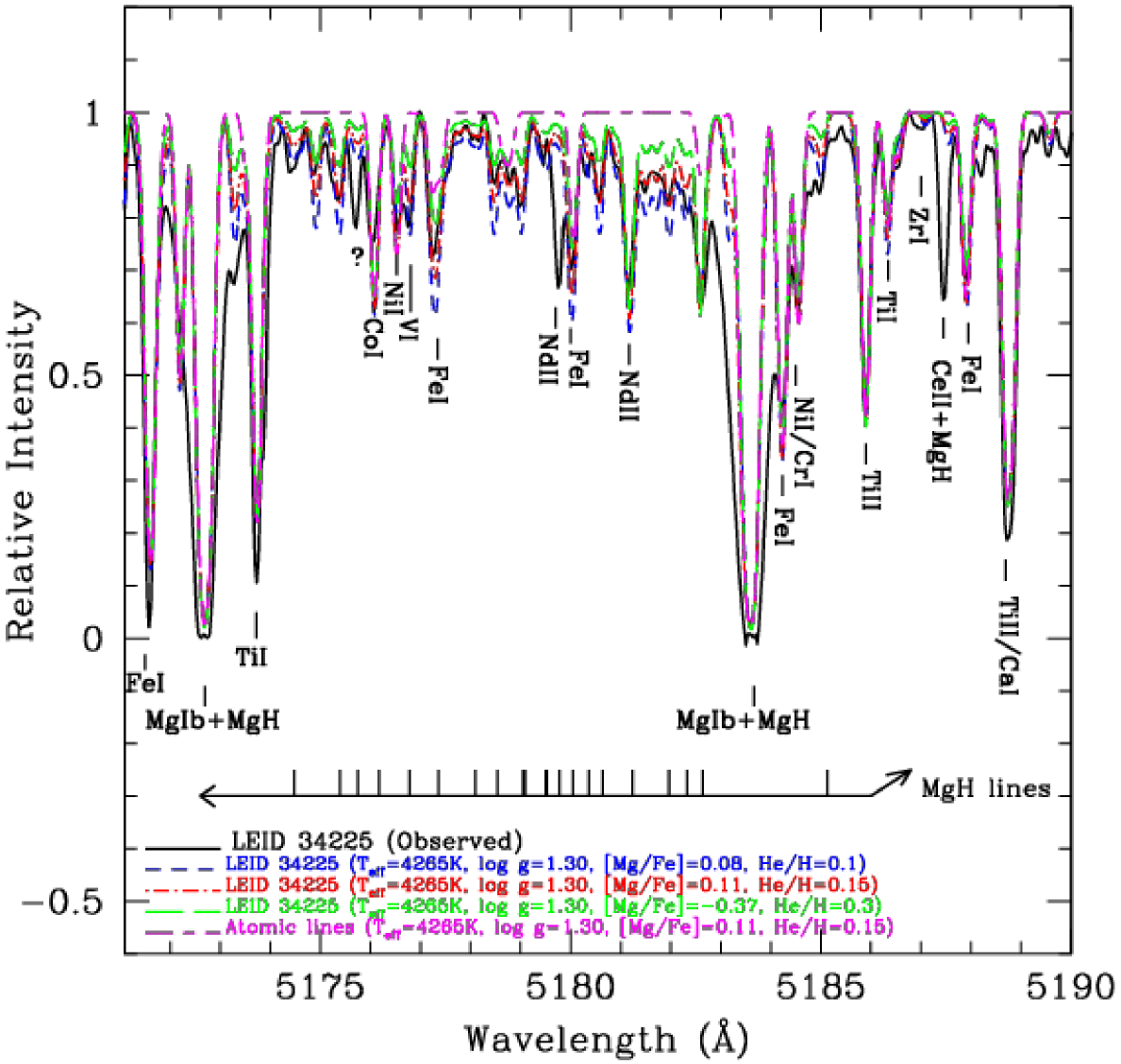}
\begin{minipage}{12cm}
\caption{Observed and the synthesized MgH bands for LEID 34225 are shown. 
The spectra synthesized for the Mg abundance derived from the Mg\,{\sc i} 
lines and the best-fit value of He/H ratio are shown by the red 
dashed-dotted line. The synthesis for the two value of the He/H are also shown.
\label{fig:2}
}
\end{minipage}
\end{figure}

From the studies of \citet{norris95}, the average
[Mg/Fe] for the red giants of $\omega$ Cen is about $+$0.4 dex over
a metallicity range: [Fe/H] = $-$2.0 to $-$0.7. Hence, in our
synthesis the [Mg/Fe] = $+$0.4 dex was initially adopted. Since
the subordinate lines of the MgH band at about 5167\AA\ are
blended with the saturated Mg b lines, the subordinate lines of
the MgH band in the wavelength window 5120--5160\AA\ were
given more weight in our synthesis. The best fit of the spectrum
synthesized for the adopted stellar parameters to the observed
was obtained by adjusting the Mg abundance, and therefore
estimating the Mg abundance for one of the program stars 
LEID\,34225 (see Fig.\,\ref{fig:2}, for example). 
Note that the derived Mg abundances are in
excellent agreement with the two common stars in the 
\citet{norris95} study. The adopted stellar parameters and the
derived Mg abundances for the program stars (first, second,
and third groups) are given in Table\,\ref{tab:1}. For all the normal
first and third group stars, our derived Mg abundances (mean
[Mg/Fe]$\sim$ $+$0.3 dex) for their adopted stellar parameters were as
expected for the red giants of $\omega$ Cen, with just four exceptions.
These four exceptions are 73170 and 178243 from the first
group and 262788 and 193804 from the third group, which were
identified with the weak/absent MgH bands in their observed
spectra. The Mg abundances derived for these four giants are
much lower than that expected (see Table\,\ref{tab:1}).

\begin{figure}[t!]
\centering
\includegraphics[scale=0.5]{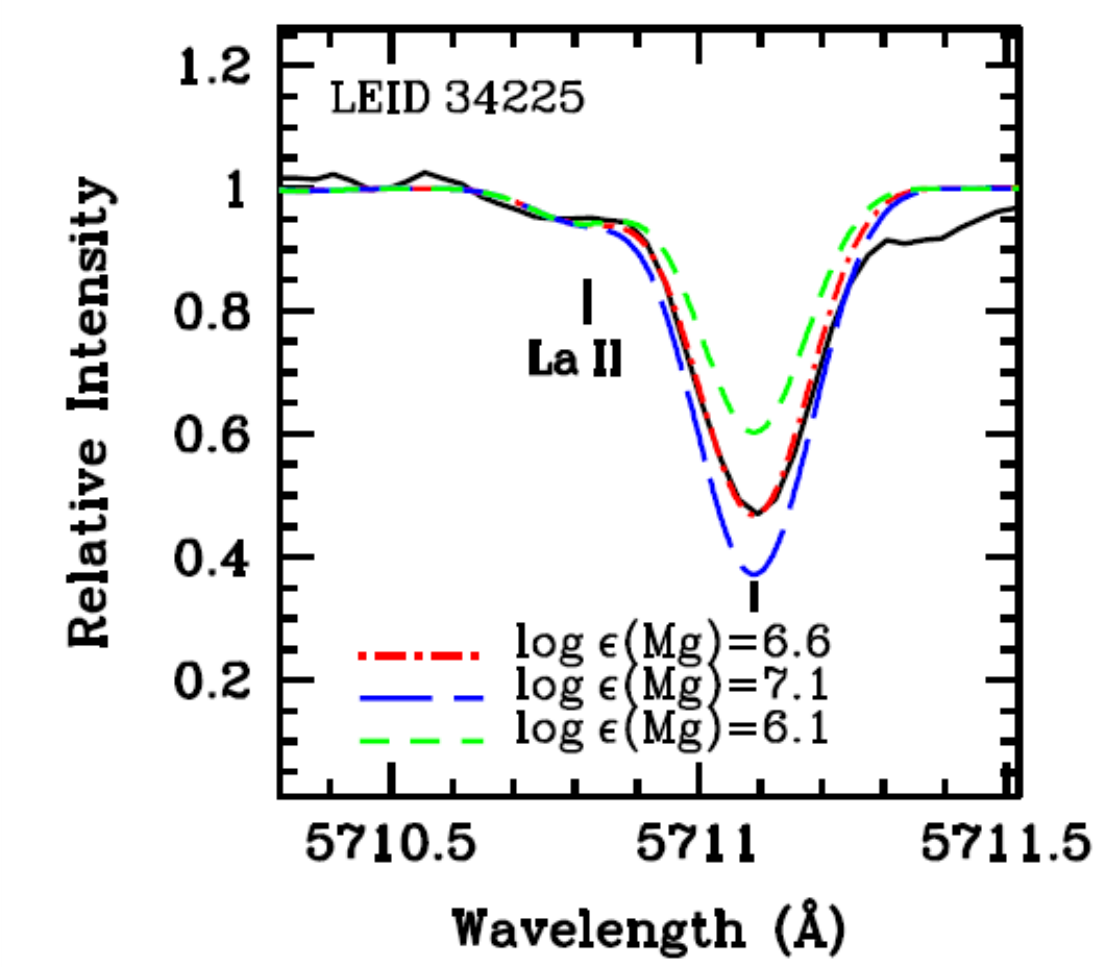}
\begin{minipage}{12cm}
\caption{The Mg abundance derived from the $\lambda\lambda$5711\AA\
Mg\,{\sc i} lines for the program stars.
The best fit synthesis is shown with the red-dash dotted line.
\label{fig:3}
}
\end{minipage}
\end{figure}

The high-resolution optical spectra obtained from 
SALT, South African Astronomical Observatory (SAAO) 
were obtained for two of 
the four metal-rich, mildly hydrogen-poor or helium-enhanced
giants, LEID 39048 (73170) and LEID 34225 (262788), 
along with their comparison normal (hydrogen-rich) 
giants, LEID 61067 (172980) and LEID 32169 (251701).
The stellar parameters and the elemental abundances for the 
program stars were derived from the equivalent widths measured from these
high-resolution spectra. The strengths of the MgH bands in the 
spectra of these program stars were analyzed with these 
derived stellar parameters. The
observed spectra of the sample (hydrogen-poor) stars 
(LEID 39048 and LEID 34225) show weaker MgH bands,
unlike in the spectra of the normal comparison giants 
(LEID 61067 and LEID 32169), as observed in the low-resolution 
spectral studies \citep{hema14}. 
The accurate Mg abundances were derived from the weaker atomic 
Mg\,{\sc i} lines (see Fig.\,\ref{fig:3}). Note that the Mg\,b lines are very strong 
and are saturated in the spectra of our program stars, 
and hence these lines are not used for estimating the 
Mg abundance from the Mg I line or MgH band. 
And, the Mg abundance from the MgH band is derived using
the clean subordinate lines of the MgH molecular band
that are not blended with the strong Mg b lines.
The spectra of the program stars were synthesized using their
derived stellar parameters and the elemental abundances as
discussed in Section\,3 of \citet{hema18}. 

Uncertainty on the $T_{\rm eff}$ and $\xi_{t}$ is 
estimated by changing the $T_{\rm eff}$  
in steps of 25 K and $\xi_{t}$ in steps of 0.05 km s$^{-1}$. The change
in $T_{\rm eff}$ and $\xi_{t}$ and the corresponding deviations in abundance,
from the zero slope abundance, of about 1$\sigma$ error, is obtained.
This change is adopted as the uncertainty on these parameters.
The adopted $\Delta$$T_{\rm eff}$=$\pm$50 K and 
$\Delta\xi_{t}$=$\pm$0.1 km\,s$^{-1}$ 
(see Fig.\,2 of \citet{hema18}). The uncertainty on log g 
is the standard deviation
from the mean value of the log g determined from different
species, which is about $\pm$0.1 (cgs units). 
The rms error due to these parameters along with the
standard deviation in the abundances due to line-to-line scatter
is the uncertainty adopted for the metallicity and Mg abundance 
from Mg\,{\sc i} and MgH band 
(please see Table\,5 of \citet{hema18}.

For the first group stars the best fit synthesis for the MgH bands 
were obtained for the Mg abundance derived from the Mg\,{\sc i} lines
within the uncertainties. But 
for the third group stars, LEID 39048 and LEID 34225, the 
Mg abundance for the best fit synthesis of MgH bands 
requires about 0.4 dex less Mg abundance than that 
derived from the Mg\,{\sc i} lines. This is perfectly in line with 
our studies \citet{hema14}. Further, the ATLAS12 opacity-sampling 
stellar model atmospheres that were iterated and generated 
for different He/H ratios were used \citet{kurucz14} 
to conduct the detailed abundance 
analysis, and also to retaliate the difference in Mg abundance 
from Mg\,{\sc i} and MgH lines by changing the He/H ratios \citep{hema20}. 
For the program stars, LEID\,39048 and LEID\,34225, the He/H ratio 
derived is about 0.2 and 0.15, respectively. For the detailed results 
of abundance analysis for the corresponding He/H ratios, please see 
\citet{hema20}.   

\section{Discussion}

Since, hydrogen and helium are the most abundant elements in the
stars' atmospheres, it is almost impossible to measure their 
accurate abundances. Only in the hotter stars, where helium 
transitions are seen, the helium abundance can be estimated.   
But, in cool stars, there are no direct measurements of helium. 
For Sun, the helium is estimated by sampling the solar wind. 
Primordial He-abundance and that from local H II regions and 
B-type stars set the floor/ceiling for He-abundance. 
Measurements of He/H ratio in cool stars are essential to 
derive accurate elemental abundances including the He-abundance.
For cool stars, elemental abundances are derived by adopting 
a "standard" He/H ratio of 0.1. 

By \citet{sumangala11},  an LTE abundance analysis conducted for 
SAO 40039, a warm post-AGB star whose spectrum is known to show 
surprisingly strong He\,{\sc i} lines for its effective temperature.
On the assumption that the He\,{\sc i} lines are of photospheric 
and not chromospheric origin, a He/H ratio of approximately unity 
was found by imposing the condition that the adopted He/H ratio 
of the model atmosphere must equal the ratio derived from
the observed He\,{\sc i} triplet lines at 5876, 4471, and 4713\AA\, 
and singlet lines at 4922 and 5015\AA.

There are stars that are severely deficient in hydrogen which 
are called the R Coronae Borealis stars, with the related groups
the extreme helium (EHe) stars and the hydrogen-deficient carbon 
(HdC) stars. These stars are having very weak/absent hydrogen-Balmer 
lines or the presence of He lines in hotter EHe stars. Hence, the 
helium abundance could be measured from these hot EHe stars.

But for cooler stars, where neither the He lines could be seen 
nor the small changes in hydrogen abundance could be measured from the 
hydrogen-Balmer lines as these are very strong and saturated. 

Hence, the method of using Mg\,{\sc i} lines and the 
MgH molecular band for measuring even smallest change in the 
He-abundance in the spectrum of cool stars is a very useful and
robust method. For the respective He/H ratios of the He-enhanced 
stars, the elemental abundances are also affected. Please see 
discussion of \citet{hema18, hema20}. 
Our study \citet{hema14, hema18, hema20} in the giants of $\omega$ Cen is 
the first ever spectroscopic confirmation of He-enhancement.  
The abundance analysis conducted by deriving the actual He/H ratio, 
not only provide the solutions to abundance anomalies observed in stars of 
the globular cluster but also for the peculiar stars of the Galactic field. 
Many normal stars are studied with the assumption of standard 
He/H ratio of 0.1, which may not be the actual value. 
Hence, more and more studies
for deriving the actual He/H ratio and in different stellar groups are 
much needed to address the He-enhancement, the abundance anomalies and the
peculiarities, to understand their origin and evolution.

\section{Conclusion}

The low-resolution spectroscopic survey conducted by \citet{hema14}
resulted in the discovery of four mild H-deficient/He-enhanced giants
in the Galactic globular cluster $\omega$ Cen. 
For two out of four giants along with their comparison stars the 
high-resolution spectra were obtained from the SALT. Using their 
high-resolution spectra and the ATLAS12 model atmospheres 
iterated for different He/H ratios, the detailed abundance analysis
were carried out \citep{hema18}. Deriving the accurate 
Mg abundances, the MgH bands 
were synthesized with model atmospheres with different He/H ratios. 
For the two He-enhanced giants: LEID 39048 and LEID 34225, 
the same Mg abundances from Mg\,{\sc i} and MgH bands were 
obtained for the He/H ratios of 0.2 and 0.15, 
respectively \citep{hema20}. And, the 
comparison stars, the derived He/H ratios were 0.1, as expected. 
This is the first ever spectroscopic confirmation for the presence
of He-enhanced giants in $\omega$ Cen. The approach of 
identifying the He-enhanced stars with the difference 
in Mg abundance from Mg\,{\sc i} and MgH band is very novel and 
robust method. Our studies bridges the evolutionary 
track of the metal-rich, He-rich population of
$\omega$ Cen.

And it is also essential to explore the origin/processes 
responsible for the He-enhancement. As said, "In light of the 
impossibility of direct detection of He-lines in photospheric 
spectra of cool stars, one is led to wonder if mildly He-rich 
cool stars exist, how they might be detected, and how they 
might arise (diffusion, internal nucleosynthesis
and mixing, binary interactions, etc?)" by \citet{lambert96}.

\begin{acknowledgments}
BPH acknowledges the Women Scientist Scheme (WOS), Department of Science and Technology (DST), India, for support through Grant: DST/WOS-A/PM-1/2020, and GP acknowledges the Science and Engineering Research Board (SERB), DST, India for support through Grant: CRG/2021/000108.
\end{acknowledgments}

\begin{furtherinformation}

\begin{orcids}
\orcid{0000-0002-0160-934X}{Banagere Prakash}{Hema}
\orcid{0000-0001-5812-1516}{Gajendra}{Pandey}
\end{orcids}

\begin{authorcontributions}
This work is part of a collective effort with contributions from all the co-authors.
\end{authorcontributions}

\begin{conflictsofinterest}
The authors declare no conflict of interest.
\end{conflictsofinterest}

\end{furtherinformation}

\bibliographystyle{bullsrsl-en}

\bibliography{S05-CT04_HemaB}

\end{document}